# Automation of Construction Quantity Take-off in a BIM Model

Hosein Taghaddos[1], Ali Mashayekhi[2], Behnam Sherafat[3]

[1] Hosein Taghaddos, PhD, PEng, Assistant Professor, School of Civil Engineering, College of Engineering, University of Tehran, Tehran, Iran, Email: htaghaddos@ut.ac.ir

[2] Ali Mashayekhi, MSc Student of Construction Engineering & Management, School of Civil Engineering, College of Engineering, University of Tehran, Tehran, Iran, Email: ali.mashayekhi@ut.ac.ir

[3] Behnam Sherafat, MSc Student of Construction Engineering & Management, School of Civil Engineering, College of Engineering, University of Tehran, Tehran, Iran, Email: sherafat@ut.ac.ir

**ABSTRACT**

Building Information Modeling (BIM) is a major upheaval in construction industry. Although BIM advantages in construction management has been proved in many papers reviewed, there are still many limitations that inhibit organizations to use BIM models efficiently. However, advancement of Application Programming Interface (API) automation in the recent years has facilitated employing BIM in construction industry. This research contributes to the state of practice in construction management by developing API codes to automate estimation of construction. The developed API automatically filters items related to a particular discipline within a particular work area and facilitate systematical quantity take off in different work areas . This allows the planner to define proper 3D work areas and estimate the required materials\man-hours in different work areas throughout the project. The results of the developed automated approach are compared with hand calculated results as well as the results calculated from the BIM software interface in special cases. Finally, the research has been validated by a case study of an actual petrochemical project.

## 1-Introduction

The importance and complexity of industrial projects makes the project managers to go through the economically efficient construction methods. Nowadays, with the increasing application of Building Information Modeling (BIM) in construction, project managers, contractors and designers willing to automate various processes in construction including estimating required materials or man-hours for different disciplines. This paper introduces a generic approach to automate the process of quantity take-off using a BIM model.

Automation in construction is related to both construction management and building science of civil engineering discipline and Information Technology (IT) discipline. Currently, IT approach to solve engineering issues in Architecture Engineering Construction (AEC) offices, is really popular where it is used to facilitate engineering calculations, construction and project management, planning and control. BIM tools as a powerful software conduct engineers to visualize steps of projects in AEC offices and to prevent rework and to reduce unexpected faults (Sacks et al. 2004).

Although BIM is capable of providing quantity take off tables, BIM tools cannot manipulate that data. This manipulation usually done with other types of software (Sattineni and Bradford II 2011). Data is usually transferred between the BIM and cost estimation software in two different ways: 1) two of each software use the same proprietary format and data transfer is done without loss of data; 2) each software use different proprietary formats and first of all the data of each software is converted to a third, common format, usually the Industry Foundation Classes (IFC). IFC [2] is an ad-hoc standard data structure for classification of AEC data that its function of providing a wide range of applications is underestimated due to its weakness of offering a vehicle for the exchange of data without data loss.

**2-Research background**

Whereas we all know that the basic objective of BIM usage is to predict future of project, lack of studies about correlation between project quantification and estimation versus with BIM tools, depict that BIM approach to solve this scope of project's problems is not that much extensive yet. This is because of many limitations that inhibit organizations to use BIM models efficiently. However, advancement of Application Programming Interface (API) automation in the recent years has facilitated employing BIM in construction industry.

However quantification of work packages like excavation could not be estimated by BIM, but we can argue if this quantities worth to be estimated specifically (Firat et al. 2010). This automation in estimation does not mean that there is no need to an engineer anymore, but an expert who ship information and compile them is still required [3]. A case study on an industrial project tried to integrate CAD-based design with schedule and cost data. This was part of the research efforts to develop and improve CAD or BIM-based cost estimation (Staub et al. 1998). Another research tried to solve cost estimation problems in design process by utility of neural network methodology (Günaydın and Zeynep Doğan 2004). Efforts continued by Cost estimation of structural skeleton using an interactive automation algorithm (Jadid and Idrees 2007) and developing a frame work for BIM-based construction cost estimating software (Ma et al. 2010). A comparative study of software that are commercially known as BIM tools that are suitable for cost estimation, and a survey of practical changes incurred by the adoption of such software in a construction organization (Forgues et al. 2012).

Most BIM tools are able to prepare a preliminary quantity take off, but this service actually depends on model to have required properties. Even if the BIM tool prepare the quantity of an element, these applications tend to lack the function to perform cost estimation, which is usually done using different software. The data exchange between cost estimator software and BIM tools is often compile via IFC. The IFC lack of lossless data exchange with each export-import process was proved and this weakness could lead us to incorrect quantities and ultimately not real cost estimation (Zhiliang 2011). This kind of issue is then solved through developing takeoff methods that provide instructor for architects, engineers and contractors on how to design, how to compile data and how to quantify the model (Kim et al. 2009). Utilizing BIM software as a best practice for quantity take off has lack of database management system and filtering issues of desired discipline to be quantified. This problem have been solved by filtering methods like or/and condition filtering in NavisWorks by using a database management system (Ali et al. 2015).

**3-Methodology**

According to research objective of finding the volumes of the objects from a BIM model, we are going to calculate other parameters like the weight of the object, cost of construction of the object and etc.

**3-1-NavisWorks Quantity Take-Off**

In order to take off a quantity from a BIM-based model, employing quantity take off features of the BIM software (e.g. Navisworks Simulate) is sometimes an option. However, o the main weakness of such software tool is that this feature is feasible if all properties of elements to be estimated model follow specific formats, called MasterFormat or UniFormat.

**3-2-Automation of Quantity Take off**

The other alternative method to estimate volumes of the model's elements in the BIM model is to employ boundary boxes around each construction element using API. Although this method has some weaknesses that we mention in result section, but it could be an appropriate approach to achieve a more accurate quantity take-off. For the objects that their boundary box's volume does not fit them, we offered a database-based solution using the following steps:

1) First different filters must be defined for desired project scope to be quantified (e.g. structural steel scope) in the BIM model. Some flexible API codes are defined previously to define different combinations of the Boolean or/and conditions, save them in a database, and group the items in the model based on these filter criteria (Ali et al. 2015).
2) Next extracted data from BIM model is stored and managed properly to a database by an API code. This data is stored and queried in the database to obtain volume, weight or cost of the desired project scope.

### 3-3- System Environment

The following software is employed to estimate the volume of the objects:

1) Autodesk Navisworks Simulate 2015: For BIM modeling

2) Microsoft Access 2010: For saving extracted data in a database format

3) Autodesk AutoCAD 2015: To draw Boundary boxes to be appended on desired elements in Navisworks.

4) Microsoft Visual Studio 2013: For writing the APIs to extract data (i.e. coordinates and volume of boundary box) from Navisworks Simulate and the to generate boundary boxes in AutoCAD

### 3-4-Quantity take-off API

This API, developed in Visual Studio 2013 environment in VB language, connects the Navisworks software to Microsoft Access and extracts coordinates of each construction element. Then it calculates the volume of each boundary box around the items (i.e. objects) in the model and then fills data in a database. This API gets the minimum and maximum points of objects, containing X, Y and Z coordinates, and estimates their volume as equation 1.

*Bounding Box Volume* $= L_x * L_y * L_z$    Where    $L_x = X_{Max} - X_{Min}$    (1)

Where: $X_{Min}, X_{Max}$, $Y_{Min}$, $Y_{Max}$, $Z_{Min}$ and $Z_{Max}$ are coordinates of minimum and maximum points (i.e. two corners) of boundary box around the item.

### 3-5-Boundary box generation API

This API, developed in Visual Studio 2013 environment in VB language, reads data from Microsoft Access and draws boundary boxes of elements in AutoCAD automatically.

### 3-6-Different element shapes issue

As a construction project usually contains elements such as cube, cylinder, wide flange column and their rotated conditions, we generated these elements in AutoCAD and appended them to Navisworks Simulate. Then the Quantity take-off API code was run to extract coordinates and the Boundary box generation API is run to draw their boundary boxes in AutoCAD. If the model element has a regular rectangular box or cubic shape, the extracted boundary box will exactly fit the object and the estimated volume is accurate. In a case that boundary box of an element does not fit its actual shape (e.g. cases shown in Figure 1), we offer a database-based solution for various types of objects.

*Cylinder*: Although boundary box of this element will not fit the cylinder, but the diameter of cylinder can be calculated from its boundary box (Figure 1-a). The diameter of cylinder equals to the two equal dimensions of the boundary box and the third dimension would be the height or length of the cylinder. The volume of such cylindrical element can be calculated as Equation 2.

$$\text{Volume} = \frac{\pi D^2}{4} \times H \tag{2}$$

Where D and H are diameter and height (or length) of a cylinder, respectively.

***Irregular Elements***: Boundary box of Irregular elements such as wide flange columns or I-beams do not fit the actual element due to their special configuration (Figure 1-c). For these particular elements, we populated a table in the database with name and section area of such irregular elements. By extracting the the name of the element from the model using the developed API, the cross section area of these elements can be calculated by a simple cross reference query. The length of element in most cases can be estimated as the maximum length of boundary box. Thus, the volume of such elements would be calculated based on Equation 3.

$$\text{Volume} = Ar \times L_{Max} \tag{3}$$

Where Ar is the cross section area and $L_{Max}$ is the maximum length of boundary box, respectively.

***Rotated Elements***: For elements that are not align x, y, or z axis, the boundary box of element would be much larger of actual element due to deviation of these elements from axes (Figure 1-b). We can overcome this issue by trigonometry calculation based on Equation 4.

$$\text{Volume} = Ar \times \sqrt{L_x^2 + L_y^2 + L_z^2} \tag{4}$$

Where $L_x$, $L_y$, $L_z$ are dimensions of the boundary box align the axes, and Ar is the cross section area calculated as discussed above.

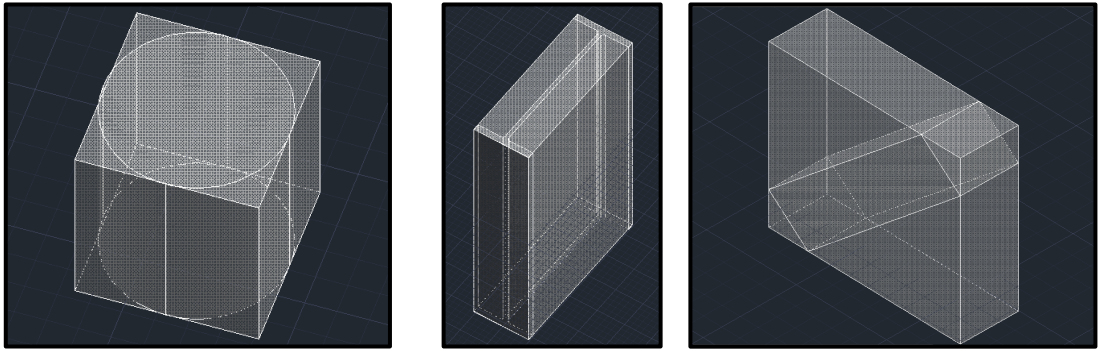

(a) Cylinder Elements    (b) Irregular Elements    (c) Rotated Elements

**Figure 1. Boundary boxes not fitting a modeling element**

**4-Results**

Figure 2 depicts part of a petrochemical project as the used case study to validate developed system. This project is designed as modular construction, where these modules can be filtered and grouped by different discipline (e.g. structural steel, piping) automatically using developed API, and the database system. This API also extracts coordinates and properties (e.g. name) of each model element to the Access database. The other API developed for AutoCAD generates rectangular boundary boxes based on coordinate's data in the database. This AutoCAD file containing boundary boxes of elements is appended to the Navisworks model for the sake of visualization and validation.

The selected module weight is *142* tones based on project reports and have two major discipline: Structural steel and Pipe lines. From project data we know that structural steel weight is about *30%* of selected module weight. Our automated database system will first filter this disciplines and then we run our code. Eventually we can compare and validate our results with project known data.

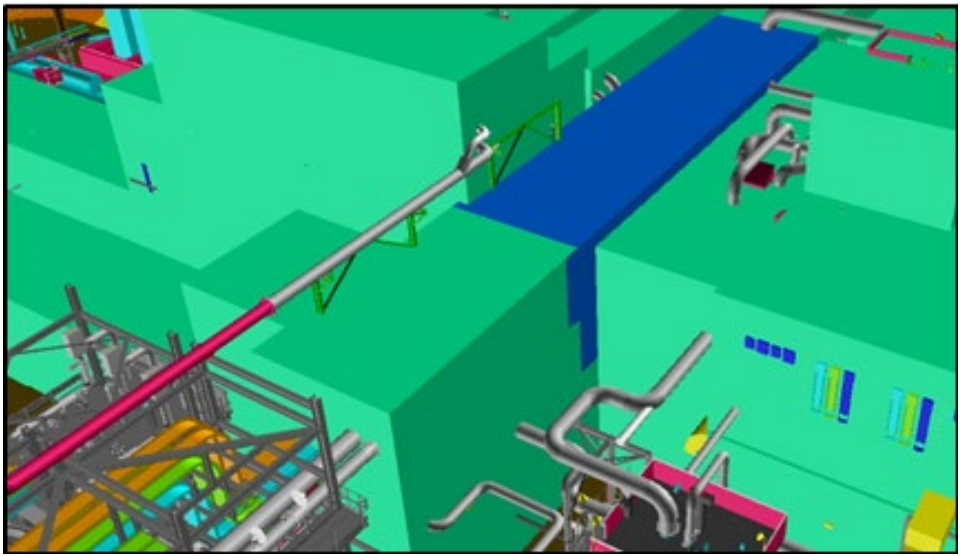

**Figure 2. Selected module (in blue) of the case study of petrochemical plant**

Figure 3 (a**)** shows all the elements of the filtered module highlighted in Figure 2. Figure 3 (b) illustrates the structural steel discipline of selected module and Figure 3 (c) depict piping discipline of this module. Figure 3 (d) demonstrates the boundary boxes of piping discipline in AutoCAD, which is then appended to Navisworks. Usage of AutoCAD in generating boundary boxes is because Navisworks software is only a reviewer application and cannot drawing boundary box objects. Figure 4 shows that structural steel elements and their boundary boxes fit well to each other. Table 1 shows how this system calculates the volume and weight of structural steel members. Section W310×118 is used for vertical elements and W310×79 is used for horizontal ones.

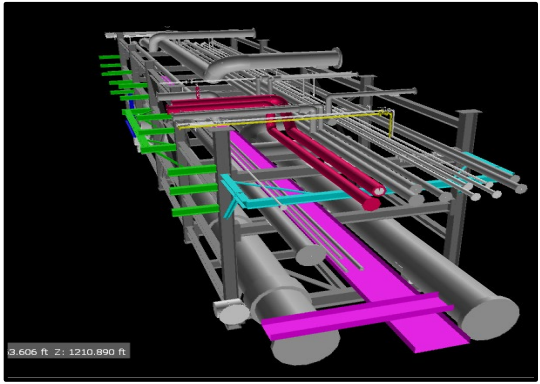 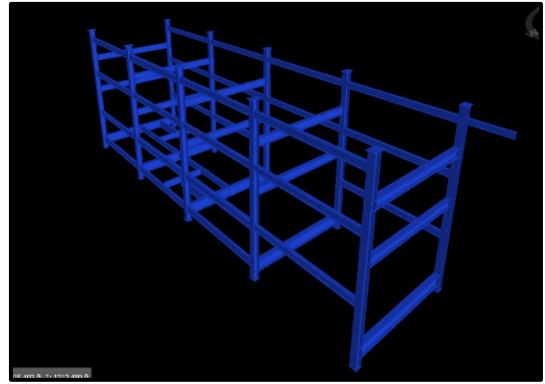

**(a) Total elements of selected module (b) Structural steel elements of the module**

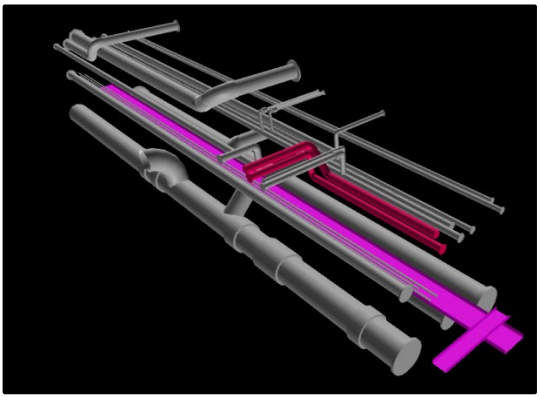 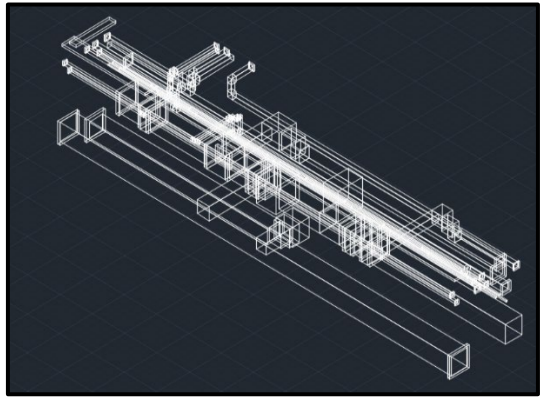

**(c) Piping elements of the module     (d) Piping boundary boxes in AutoCAD**

**Figure 3. Structural steel and piping boundary boxes for a module**

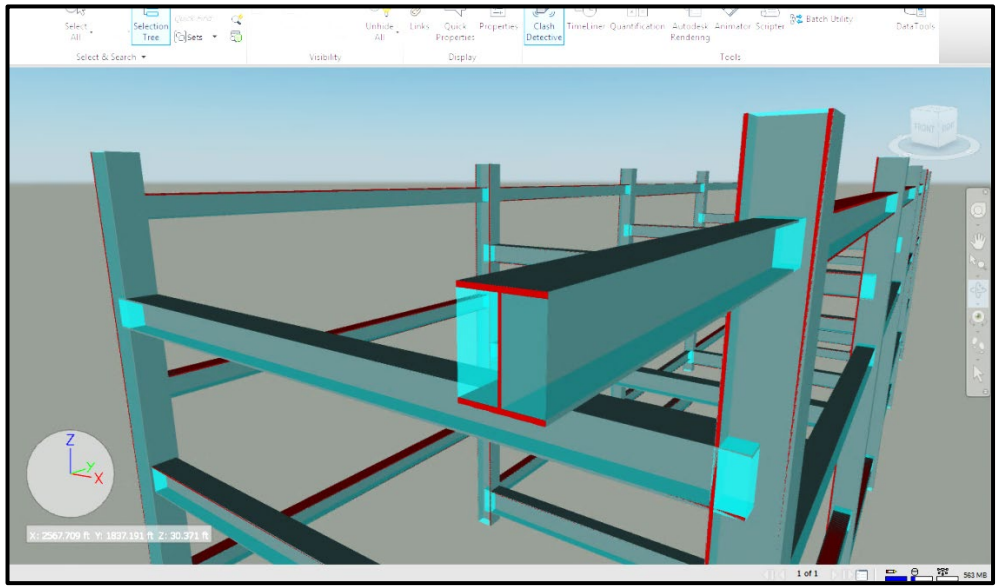

**Figure 4. Structural steel element and their boundary boxes in Navisworks**

**Table 1. Automated estimation of structural steel discipline**

| Row Number | Element Type | Lx (m) | Ly (m) | Lz (m) | Height (m) | Area (m2) | Volume (m3) | Density (kg/m) | Total Mass (kg) |
|---|---|---|---|---|---|---|---|---|---|
| 1 | W310×79 | 6 | 0.3 | 0.3 | 6.000 | 0.0101 | 0.0606 | 80.661 | 483.967 |
| ⋮ | ⋮ | ⋮ | ⋮ | ⋮ | ⋮ | ⋮ | ⋮ | ⋮ | ⋮ |
| 8 | W310×118 | 0.3 | 0.3 | 6.9 | 6.950 | 0.0151 | 0.10494 | 119.68 | 831.775 |
| ⋮ | ⋮ | ⋮ | ⋮ | ⋮ | ⋮ | ⋮ | ⋮ | ⋮ | ⋮ |
| 60 | W310×79 | 0.3 | 6 | 0.3 | 5.991 | 0.0101 | 0.06051 | 80.661 | 483.208 |
| 61 | W310×79 | 0.2 | 6 | 0.3 | 6.000 | 0.0101 | 0.0606 | 80.661 | 483.965 |
|  |  |  |  |  |  | **Total Volume (m3)** | *3.9568* | **Total Mass (ton)** | *31.5363* |

The calculated total weight of structural steel members of selected module is about 30 tones, which is very close to the actual structural steel weight of the module, and about 20% of the total actual weight of the module.

Figure 5 depicts piping discipline of the selected module and boundary boxes of these elements in Navisworks. Table 2 shows the parameters of a pipe line extracted from the model and its associated information. Table 3 validates the results of API-based automated estimation data for the same pipe and the proposed method. This table shows that the difference of the calculated weight and the model data of this selected pipe line are negligible.

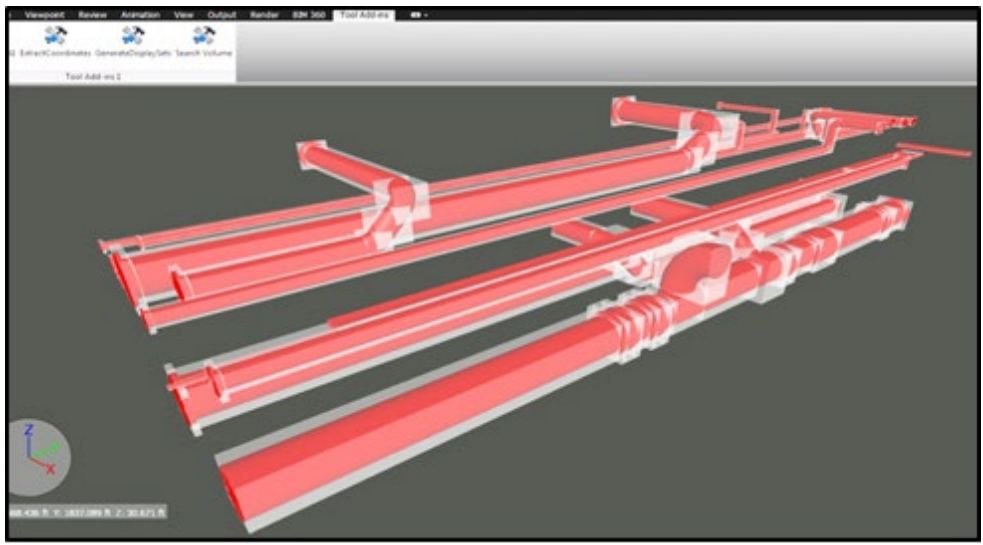

**Figure 5. Piping filter and its elements boundary boxes in Navisworks**

**Table 2. Known data of a selected pipe line of the selected module**

| NPD | Outer Diameter(m) | Weight(ton) | Length(m) | Pipe Type |
|---|---|---|---|---|
| 24 | 0.6096 | 5.3056021 | 14.94403 | Sch/thk:S-60 |
| Thickness(m) | Inner Diameter(m) | Area(m2) | Volume(m3) | Density(ton/m3) |
| 0.02461 | 0.56038 | 0.04520533 | 0.675549901 | 7.853753057 |

**Table 3. The proposed automated estimation of selected pipe line**

| Lx(m) | Ly(m) | Outer Diameter(m) | Area(m2) | Weight(tom) |
|---|---|---|---|---|
| 0.60957497 | 14.94404489 | 0.609574974 | 0.045203402 | 5.305380407 |
| Lz(m) | Thickness(m) | Inner Diameter(m) | Volume(m3) | Differences |
| 0.60957497 | 0.02461 | 0.560354974 | 0.675521673 | **-0.0042%** |

**5-Conclusion**

This paper focuses on developing an automated system to estimate the material. Although some features to assist the modeler in quantity take off are embedded in some BIM applications (e.g. Navisworks), using this feature is not always feasible due to lack of the required properties in the model. The proposed system is validated in an actual case study of an industrial construction.

Estimating volume and weight can results in man-hour estimating by having the productivity data and consequently cost estimation. This automated approach can assist the project team in bidding, procurement, planning purposes in a rational manner.

**6- Further Research**

Our future study concentrates on methods that extracts the points on the elements from the BIM model and calculates the object perimeter to fit more accurate boundary shape. One of these methods is K-DOP method which we would like to investigate in the future work to make our API code more practical and accurate.

# 7-References